\documentclass[twocolumn,showpacs,preprintnumbers,amsmath,amssymb,prb]{revtex4}

\usepackage{graphicx}
\usepackage{dcolumn}
\usepackage{bm}
\usepackage{amsfonts}
\usepackage{latexsym}
\usepackage{amsmath}
\usepackage{amsbsy}
\usepackage{amssymb}

\newcommand{\HoTi}{Ho$_2$Ti$_2$O$_7$ }
\newcommand{\DyTi}{Dy$_2$Ti$_2$O$_7$ }
\newcommand{\DyTins}{Dy$_2$Ti$_2$O$_7$}
\newcommand{\HoTins}{Ho$_2$Ti$_2$O$_7$}
\newcommand{\HoGans}{Ho$_2$GaSbO$_7$}
\newcommand{\degrees}{\ensuremath{^\circ}}

\newcommand{\HoSnns}{Ho$_2$Sn$_2$O$_7$}

\begin{document}

\title{Neutron scattering studies of the spin ices \HoTi and \DyTi in
applied magnetic field}

\author{T. Fennell}
\altaffiliation[Now at ]{ESRF, 6 Rue Jules Horowitz, BP220, 38043
Grenoble Cedex9, France}\affiliation{The Royal Institution of
Great Britain, 21 Albemarle Street, London, W1S 4BS, United
Kingdom}

\author{O. A. Petrenko}\altaffiliation[Now at ]{Department of Physics,
University of Warwick, Coventry, CV4 7AL, United Kingdom} \affiliation{ISIS
Facility, Rutherford-Appleton Laboratory, Chilton, Didcot, OX11 0QX, United
Kingdom}

\author{J. S. Gardner}\affiliation{Physics Department, Brookhaven National Laboratory, Upton, NY11973-5000, \& \\
NIST Center for Neutron Research, National Institute of Standards
and Technology, Gaithersburg, Maryland, 20899-8562, USA}

\author{S. T. Bramwell}\email{s.t.bramwell@ucl.ac.uk}
\affiliation{Department of Chemistry, University College London,
20 Gordon Street, London, WC1H~0AJ, United Kingdom}

\author{B. F{\aa}k}\altaffiliation[Permanent address ]{CEA Grenoble, DRFMC/SPSMS, 38054,
Grenoble Cedex, France} \affiliation{ISIS Facility,
Rutherford-Appleton Laboratory, Chilton, Didcot, OX11 0QX, United Kingdom}

\author{B. Ouladdiaf} \affiliation{Institut Laue-Langevin, 6, rue Jules
Horowitz, BP 156-38042, Grenoble, Cedex 9, France}

\date{\today}

\begin{abstract}

Neutron diffraction has been used to investigate the magnetic
correlations in single crystals of the spin ice materials \HoTi
and \DyTi in an external magnetic field applied along either the
$[001]$ or $[1\bar{1}0]$ crystallographic directions.  With the
field applied along $[001]$ a long range ordered groundstate is
selected from the spin ice manifold. With the field applied along
$[1\bar{1}0]$ the experiments show that the spin system is
separated into parallel ($\alpha$) and perpendicular ($\beta$)
chains with respect to the field.  This leads to partial ordering
and the appearance of quasi-one-dimensional magnetic structures.
In both field orientations this frustrated spin system is defined
by the appearance of metastable states, magnetization plateaux and
unusually slow, field regulated dynamics.

\end{abstract}

\pacs{75.25.+z, 75.40.Cx, 75.50.-y, 75.60.Ej}

\maketitle

\section{\label{intro}Introduction}

The spin ices~\cite{prl1,cult}, such as
\HoTins~\cite{prl1,newprl}, \DyTins~\cite{art} and
\HoSnns~\cite{stannate} are frustrated Ising magnets.  The rare
earth atoms are located on a pyrochlore lattice, a face centered
cubic (fcc) lattice with tetrahedral basis, creating a network of
corner-linked tetrahedra (see Fig.~\ref{q0tet}).  In these systems
the crystal field creates a very strong anisotropy estimated at
$\approx 300$ K~\cite{rosenkranz} which constrains the spins to
the local trigonal axes (members of the $\langle 111 \rangle$
direction set).  The spins are therefore Ising-like but
non-collinear: they point either into or out of the elementary
tetrahedra of the pyrochlore structure~\cite{prl1,cult}. The
dominant interaction is the dipolar interaction which favors a
local arrangement of spins on a tetrahedron in which two spins
point in and two spins point out~\cite{byron}.  This is analogous
to the Bernal-Fowler ``ice rule'' for proton location in ice and
means that the spin ice has the same configurational entropy as
ice~\cite{cult,art}.  The anisotropy is so strong that the fully
saturated ferromagnetic state can be reached only in principal,
using magnetic fields well outside the range of conventional
measurements~\cite{olegmag,fukazawa}. Most of the interesting
physics happens in much lower fields, where the Zeeman energy,
which competes with relatively weak exchange and dipolar
interactions, is dominated by the $\langle 111 \rangle$
anisotropy. For instance, in polycrystalline \DyTins, about half
of the zero-point entropy missing in zero field, $(1/2) R \ln
(3/2)$, is recovered already in a field of 0.5 T~\cite{art}.

\begin{figure}
\caption{\label{q0tet} From top: groundstate of a single
tetrahedron with field (shown by arrow) applied along $[001]$;
groundstate with field applied along $[1\bar{1}0]$, the field
separates the spins into $\alpha$ (parallel to field) and $\beta$
(perpendicular to field) chains; long range ordered structure with
$\beta$ chains all parallel (called $Q=0$); $\beta$ chains
antiparallel (called $Q=X$) (both reproduced from
Ref.~\onlinecite{prl1}).}
\end{figure}

In general the spin ices have strongly anisotropic bulk
properties~\cite{cornelius,olegmag,fukazawa}.  This is unusual for
cubic systems but can be understood as another manifestation of
the strongly anisotropic nature of the individual rare earth
moments.  The application of magnetic fields at low temperature
can be used to systematically investigate the removal of the
degeneracy of the spin ice state. As a result of the anisotropy,
different degeneracy removal schemes are expected depending on the
direction in which the field is applied~\cite{lgcp}. Several
scenarios have been investigated by
experimental~\cite{art,cornelius,olegmag,fukazawa,higashinaka1,kagice1,kagice2,kagiceprl,higashinaka2,higashinaka3,hiroi,higashinaka4,aoki}
and
theoretical~\cite{lgcp,apostrophe,aswkagice,rod111,yoshida,melkorev,isakov,ruff}
methods but there have so far been only a few neutron scattering
investigations~\cite{prl1,qiu,tfthesis,spinice5}.

The basic behavior of the system can be understood in terms of the
near neighbor spin ice model, which considers near neighbor
ferromagnetic coupling between the spins~\cite{lgcp}. Application
of the field along the $[001]$ direction is expected to completely
remove the degeneracy that exists in zero field~\cite{lgcp}. All
the spins on a tetrahedron have the same relation to the applied
field, making an angle of 54\degrees~to it (see Fig.~\ref{q0tet}).
Fields up to 2 T along $[001]$ are insufficient to overcome the
anisotropy and fully align the spins. However, the field does
stabilize a ground state in which the same member of the set of
six ``two-spins-in-two-spins-out'' local groundstates is selected
on every tetrahedron. Experimentally~\cite{olegmag,fukazawa} this
is manifested, at 1.6 K, by a simple magnetization curve and a
saturation magnetization of $5.78~\mu_\mathrm{B}$ atom$^{-1}$. The
saturation magnetization is reduced from $10.0~\mu_\mathrm{B}$
atom$^{-1}$, as expected for the non-collinear structure.

The near neighbor spin ice model predicts a particularly
interesting phase transition when the field is applied along
$[001]$~\cite{lgcp}.  In zero field the model system shows no
ordering transition and the heat capacity consists of a single
Schottky anomaly due to isolated spin flips.  In small fields a
very sharp peak occurs below the Schottky anomaly, indicative of a
first order phase transition. The field-temperature phase diagram
shown in Fig.~\ref{lgcpfig}, has a line of symmetry-sustaining
first order phase transitions, terminated by a critical end point.
This is analogous to the density variation seen in a
pressure-temperature phase diagram of a liquid-gas system. The
``liquid'' and ``gas''phases of the magnetic system are
distinguished thermodynamically by their degree of magnetization,
that plays a role analogous to the fluid density. A triple point
occurs at $H = T= 0$. One aim of the work described in this paper
is to search for the critical point in \HoTins. This is described
in section~\ref{ho001ressec}.

\begin{figure}
\caption{\label{lgcpfig}The phase diagram of the near neighbor
spin ice model with field applied along $[001]$~\cite{lgcp}.  With
the field applied in this direction the model shows an unusual
liquid-gas type behavior.  A line of first order phase transitions
from the gas-like region (G) to the liquid-like region (L) is
terminated by a critical point where the transition becomes
continuous.}
\end{figure}

When the field is applied along $[1\bar{1}0]$ two spins per
tetrahedron have a component in the field direction, while two are
precisely perpendicular to the field (see Fig. \ref{q0tet}). Of
the two which have a component along the field, one is pinned
pointing ``in'' and one pointing ``out'' of the tetrahedron. The
perpendicular spins are not pinned by the field. However, in the
applied field, the ground state of the tetrahedron can still be
described by the ``ice rule''.  Therefore the unpinned spins also
form an in/out pair (see Fig.~\ref{q0tet})~\cite{prl1,cult}.  The
two groups form in-out spin chains, running either parallel or
perpendicular to the field. Each chain has a magnetic moment in
the sense of the chain, but in the near neighbor model there is no
coupling between the perpendicular chains.  We will use the
terminology of Hiroi~\cite{hiroi}, who named the chains parallel
(perpendicular) to the field $\alpha$ ($\beta$) chains.

In the neutron scattering study of Harris {\it et al.}~\cite{prl1}
a single crystal of  \HoTi was aligned with the field applied
along $[1 \bar{1} 0]$.  In zero field \HoTi has a disordered
ice-like state with zero point entropy~\cite{newprl,cornelius}. In
Ref.~\onlinecite{prl1} it was shown that the application of the
field along $[1 \bar{1} 0]$ restores order, with two coexisting
magnetic structures, denoted $Q=0$ and $Q=X$ (see
Fig.~\ref{q0tet}).  The structures are named because $Q=0$ gives
rise only to magnetic Bragg scattering at zone centers, or $Q=0$
positions, whereas $Q=X$ gives rise to Bragg scattering at both
the zone centers and the $X$ points on the zone boundaries (see
Fig.~\ref{zonesdetectors}).  The $Q=0$ structure was suggested to
have all the chains, both $\alpha$ and $\beta$, aligned parallel
with their neighbors. The moment of the structure is at $45
\degrees$ to the applied field. The $Q = X$ structure was
suggested to have alternate $\beta$ chains antiparallel, with the
net moment in the direction of the applied field.  It was found
that the $Q=0$ structure was formed at 0.3 K when a field was
applied and the $Q=X$ structure appeared as a high temperature
($\approx 1$ K) modification.  The Ho$^{3+}$ moment is severely
reduced in the $Q = X$ phase~\cite{prl1}. One of the results of
the current work is to show that the $Q=0$ and $Q=X$ structures
are in fact associated with different sets of spins: the $\alpha$
and $\beta$ chains respectively and that the formation of $Q=X$
correlations is controlled by the very slow dynamics in the
$T<0.5$ K regime (see sections~\ref{dy110ressec}
and~\ref{ho110ressec}).

In addition to making the first determination of the residual
entropy in a spin ice~\cite{art}, Ramirez and co-workers measured
the heat capacity of a \DyTi powder in fields up to 6 T. Three
small, sharp, field independent peaks were observed in finite
fields. These peaks at $T_p = 0.34$, $0.47$ and $1.12$ K were
attributed to the ordering of spins not pinned by the field. More
recently, single crystal samples of \DyTi have been studied by
bulk methods in fields applied along $[110]$~\cite{hiroi}. The
experimental results of Ref.~\onlinecite{hiroi} suggest the
formation of the $\beta$ chains, but do not infer any correlation
between the chains. It was concluded that the $\beta$ chains
behave as one dimensional Ising ferromagnets.  Most recently,
simulations of the dipolar spin ice model (that goes beyond the
near neighbor model by considering further range dipolar
couplings) have predicted that in strong fields there are two
transitions~\cite{yoshida,ruff}. At high temperature the field
induces order in the $\alpha$ chains and at a lower temperature
the $\beta$ chains undergo a phase transition into the $Q = X$
phase.

\DyTi has been studied less by neutron scattering due to the
significant absorption problem.  However, it has been shown that
at high temperatures ($T>4$ K) \DyTi is paramagnetic with
fluctuations dominated by single ion effects~\cite{qiu}. A
preliminary report of some of the results detailed below has also
been made~\cite{spinice5}.

It is now generally accepted that the spin ice materials \HoTi and
\DyTi only differ in detail; a comparison of the two should
identify generic behavior and reveal the importance of minor terms
in the spin Hamiltonian. In this work \HoTi and \DyTi have been
studied by single crystal neutron scattering with an external
magnetic field applied along the $[001]$ or $[1\bar{1}0]$ axis.
These measurements are the first detailed microscopic
investigation of the field dependent properties of \DyTi at low
temperature, and an extension of the investigation already
performed on \HoTins~\cite{prl1}.

The rest of the paper is organized as follows. Following an
experimental section (\ref{expsec}), results are given for
application of the field along $[001]$ in section~\ref{ressec001}
and discussed in section~\ref{dissec001}. Then the results for the
$[1\bar{1}0]$ orientation are given and discussed in sections
~\ref{ressec110} and~\ref{dissec110}. As yet there is no detailed
theory with which to analyze these results, so we restrict
ourselves to largely qualitative discussion and conclusions
(section~\ref{consec}).

\section{Experimental}\label{expsec}

The experiments on \HoTi and \DyTi were conducted on different
instruments at two neutron sources: \HoTi on D10 at the ILL
reactor source, and \DyTi on PRISMA at the ISIS spallation source.

All the measurements reported in this article used a single
crystal aligned with either the $[001]$ or $[1\bar{1}0]$ axis
parallel to the applied field (vertical). This allowed us to
access the scattering plane with wavevectors of the type $h,k,0$
or $h,h,l$ respectively (see Fig.~\ref{zonesdetectors} for an
illustration of these planes).

\begin{figure*}
\caption{\label{zonesdetectors}The reciprocal space of
Dy$_2$Ti$_2$O$_7$ in the $h,k,0$ (left) and $h,h,l$ (right) planes
as studied on PRISMA.  The $\mathbf{Q} = 0$ and $\mathbf{Q} = X$
positions are indicated by {\tt o} and {\tt x} respectively. In
both diagrams the area of scattering plane that lies within the
first aluminium powder line and was included in a mapping is
indicated by the curving shells.  With the field along $[001]$
($h,k,0$ scattering plane), single setting scans were measured at
intermediate field points.  In this case 16 detectors cover the
region between the two radial detector trajectories enclosing the
$h,0,0$ axis.  With the field along $[1\bar{1}0]$ ($h,h,l$
scattering plane), rocking scans were measured.  In this case 80
detector trajectories cover the region between the two radial
detector trajectories enclosing the $0,0,l$ axis.  This gives
vastly improved reciprocal space coverage compared to a single
setting scan such as that indicated about the $h,h,0$ axis, where
only 16 detector trajectories cover the same area of reciprocal
space.}
\end{figure*}

\subsection{\DyTi on PRISMA}

Two experiments were carried out on the PRISMA spectrometer at
ISIS under the same conditions. To minimize neutron absorption,
the isotopically enriched single crystal of $^{162}$\DyTins,
described in detail in Ref.~\onlinecite{series1}, was used.  The
crystal was varnished into a large copper support to ensure good
thermal contact and mounted on a dilution refrigerator insert in a
7 T cryomagnet.

The 16 detectors of PRISMA make radial trajectories across the
scattering plane~\cite{zink}.  In order to map the scattering
plane, the crystal is rotated about the vertical axis and
scattering from adjacent sectors of the scattering plane is
collected in successive settings.  To cover a small area of the
scattering plane two different types of scan can be used.  Either
a single setting of the crystal providing relatively coarse
coverage of reciprocal space, or for better resolution a rocking
scan.  In this case the 16 detector trajectories of successive
points in the rocking scan are interleaved, considerably improving
the definition of features in reciprocal space.  The scattering
planes and detector trajectories used in this study are
illustrated in Fig.~\ref{zonesdetectors}.

Where non-resolution limited features were observed they were used
to extract correlation lengths.  The resolution function as a
function of time of flight (parallel to the detector trajectory)
is a bilateral exponential function describing the pulse shape
from the moderator~\cite{hagenfunction}. Perpendicular to the
detector trajectory the resolution function is Gaussian. The
resolution functions were parameterized and interpolated by
fitting to Bragg peaks in the high field data and convolved with a
Lorentizian to fit the experimental data.

\subsection{\HoTi on D10}

\HoTi was measured on the single crystal diffractometer D10 at the
ILL. Two single crystals grown from a flux of lead fluoride were
used~\cite{wanklyn}, one for each field direction. For the $[001]$
orientation the crystal was $\approx 3\times 3 \times 3$ mm and
for the $[1\bar{1}0]$ orientation the crystal was $\approx 2\times
2 \times 2$ mm. They were of identical quality if the widths of
Bragg peaks were compared.  To simplify corrections for effects
such as demagnetization and extinction, the crystals were
sphericalized~\cite{spheric}.

Appropriately aligned crystals were fixed to oxygen free copper
pins using stycast resin and attached to a dilution refrigerator
insert in a 2.5 T vertical field cryomagnet, mounted on D10 using
a tilt stage. D10 was configured with a graphite monochromator and
a small area detector.  The wavelength was $2.56$ \AA~ and a
pyrolitic graphite filter in the incident beam was used to
suppress any $\lambda / 2$ contamination. $\omega$-scans (in which
the crystal is rotated about the vertical axis) were measured at
all positions.

The integrated intensities of nuclear and magnetic Bragg peaks
were extracted in the usual manner~\cite{garrynclive}. However,
the diffuse scattering was often too broad for the $\omega$-scan
performed. This resulted in the appearance of a flat intensity
profile, albeit elevated from the expected background, whose
intensity and width were subsequently strongly field or
temperature dependent. It was therefore attributed to a magnetic
scattering feature broader than the width of the scan. It proved
impractical to extract widths from these features so the peak
intensity was obtained to give a measure of $\chi_{\mathbf{Q}}$,
the wavevector dependent susceptibility, throughout the
experiment. The amplitude was extracted by fitting to a functional
form appropriate to the peak shape: straight lines, single
Gaussians, or triangular functions.

The Bragg scattering data was treated by refining model magnetic
structures using the CCSL~\cite{ccsl} program MAGLSQ.  During the
data analysis it became apparent that there was a significant
extinction problem.  In a subsequent experiment the larger crystal
was measured at room temperature using four circle geometry and
two wavelengths ($1.26$ and $2.36$ \AA, copper and graphite
monochromators respectively). This data was used for the
refinement of extinction parameters using the Becker and Coppens
model as implemented in the CCSL program
SFLSQ~\cite{beckerncoppens,ccsl}.  Thermal parameters derived from
a Rietveld refinement against powder diffraction data were used to
avoid correlation between the extinction and thermal
parameters~\cite{tfthesis}.  The powder data was recorded at 1.8 K
on the POLARIS medium resolution diffractometer at ISIS. The
difference in temperature of the single crystal and powder data
sets was assumed to be absorbed by the extinction parameters as
the final fit was good ($R_3 = 4 \%$~\cite{rfactor}).  The
extinction parameters allow the reproduction of nuclear
intensities so that the magnetic structure can be fitted by
adjusting only the size of the magnetic moment.

The quality of the fit of the magnetic structures decreases as the
field increases (in Fig.~\ref{d10hk0loops} it can be seen that as
the field increases, the uncertainty of the fitted moment becomes
much greater). This is attributed to a field dependent extinction
effect.  It is possible for nuclear and magnetic scattering to be
affected differently by extinction~\cite{mixed}.  Since the
magnetic structures in this work are field dependent, it is likely
that the extinction is also field dependent. As the field and
magnetic intensity increases, the extinction correction becomes
less and less applicable.  It proved impossible to overcome the
problem with the available data.

\section{Results for application of field along $[001]$}\label{ressec001}

\subsection{\DyTi}\label{dyressec001}

The sample was cooled to 0.05 K and the $h,k,0$ scattering plane
was mapped. Then the first quadrant of a hysteresis loop was
measured (up to 2 T), during which the scattering plane was mapped
at 1 and 2 T.  At intermediate fields, a single setting was
measured giving access to a small area of reciprocal space
surrounding the $h,0,0$ axis out to the $(8,0,0)$ reflection.
After returning to zero field the crystal was rotated to access
the $(4,\bar{2},0)$ peak and its field dependence was measured.

Fig.~\ref{twomaps} compares the scattering in 0 and 1 T from the
$h,k,0$ plane.  The diffuse scattering entirely disappears and is
replaced by magnetic Bragg scattering at $Q=0$ positions
consistent with the formation of the expected canted ferromagnetic
phase.  This is direct confirmation of the complete removal of
degeneracy and field induced ordering.

\begin{figure}
\caption{\label{twomaps}(Color Online) \DyTi ($B$ // [001]):
scattering intensity in the $h,k,0$ plane recorded on PRISMA at
0.05 K in zero field (top), and in 1 T (bottom).  The strong
diffuse scattering characteristic of the short range ordered spin
ice state disappears entirely, giving way to magnetic Bragg
scattering. Note that the intensity scale is cut off at 2 to
reveal the presence/absence of diffuse scattering, the maximum
intensity of a Bragg peak is much greater.}
\end{figure}

The formation of the ordered phase is also notable.
Fig.~\ref{twoloops} shows the integrated intensity of two purely
magnetic Bragg peaks. This magnetic intensity grows in a series of
steps. Successive measurements produce similar, but not identical,
order parameter curves indicating a strong history dependence.

\begin{figure}
\caption{\label{twoloops}\DyTi ($B$ // [001]): Integrated
intensity of the $(2,0,0)$ peak on the first field cycle and the
$(4,\bar{2},0)$ on the second cycle with field applied along
$[001]$ (the intensity of $(2,0,0)$ has been divided by 2).   Both
peaks are purely magnetic scattering. The history dependence and
existence of metastable states during the magnetization process is
clearly seen by the different hysteresis in the two cycles and the
plateaux. Lines are to guide the eye only, filled symbols indicate
the falling field leg.}
\end{figure}

\subsection{\HoTi}\label{ho001ressec}

In an attempt to find the anticipated liquid-gas critical
point~\cite{lgcp} a detailed survey of $B/T$-space was performed.
Field cycles up to 2.5 T were performed at 0.05, 0.15, 0.3, 0.5,
0.7 and 1.2 K.  Each was separated by warming to 1.2 K and cooling
in zero field. For this study four reciprocal lattice positions
were measured, two $Q=0$ positions ($(2,0,0)$ and $(4,2,0)$) and
two $Q=X$ positions ($(3,0,0)$ and $(1,0,0)$).

While cycling the field, magnetic scattering was found to develop
only on the $Q=0$ Bragg positions.  The loops at all temperatures
are consistent with the development and saturation of the $Q=0$
structure (Fig.~\ref{d10hk0loops}). In all cases $R_3 < 10
\%$~\cite{rfactor}. The value of the ordered moment obtained
($10.3 \pm 0.4 ~ \mu_\mathrm{B}$ atom$^{-1}$) agrees well with
that expected for Ho$^{3+}$ ($10.0~\mu_\mathrm{B}$ atom$^{-1}$).
Again it is clear that the degeneracy is completely removed by the
application of the field in this direction. Increasing hysteresis
and persistence of metastable states was observed as the
temperature was decreased below 0.5 K. At 0.05 K a remnant
magnetization is frozen into the sample on return to zero field.

\begin{figure}
\caption{\label{d10hk0loops}\HoTi ($B$ // [001]): development of
$\mathbf{Q}=0$ magnetization at different temperatures, with field
applied along $[001]$.  As the temperature decreases increasing
hysteresis and metastable states appear in the magnetization,
although the final ordered state is always the same.  Lines are to
guide the eye only, filled symbols indicate the falling field leg,
successive higher temperature loops are shifted by -1 T and + 0.5
$\mu_\mathrm{B}$ atom$^{-1}$.}
\end{figure}

Although no Bragg scattering was observed at the $Q=X$ positions,
strong diffuse scattering associated with the spin ice
state~\cite{newprl} was observed in zero field.  The intensity of
this scattering during the field cycle at 0.05 K is compared to
that of the $(2,0,0)$ in Fig.~\ref{hk0bgs}.  The diffuse
scattering persists in finite field and shows a step-like decrease
commensurate with the step-like increase in the $(2,0,0)$
scattering.  This indicates that disordered regions persist in the
sample in finite field.  The diffuse scattering disappears as the
Bragg scattering saturates in the fully ordered state. This
relationship between $Q=0$ Bragg and $Q=X$ diffuse scattering is
true at all temperatures measured, but at elevated temperatures
($T > 0.5$ K) there is no hysteresis.

\begin{figure}
\caption{\label{hk0bgs}\HoTi ($B$ // [001]): the intensity of the
diffuse scattering, at the $(1,0,0)$ position as a function of
field, at 0.05 K (top). The $(2,0,0)$ peak intensity is shown as a
comparison (it has been divided by 50).  As the fully ordered
structure develops the sample passes through a partially ordered
regime in which both Bragg scattering and diffuse scattering are
observed (between 0.3 and 0.7 T).  At the end of the field cycle
the sample is frozen in a partially ordered state with a remnant
magnetization.  As the temperature is raised above 0.5 K, the
remnant magnetization decays and the diffuse scattering
characteristic of the zero field spin ice state is reestablished
(bottom). Lines are to guide the eye only, filled symbols indicate
the falling field leg.}
\end{figure}

After the hysteresis loop at 0.05 K was completed, the temperature
was raised in zero field.  As discussed above and shown in
Fig.~\ref{d10hk0loops}, there is a remnant magnetization at the
point where the hysteresis loop returns to zero field.  By
comparing the Bragg and diffuse scattering intensities during the
temperature rise, it can be seen that the frozen long range order
created by applying high fields at low temperatures decays and the
short range ordered spin ice state is re-established as the system
is warmed up (see Fig.~\ref{hk0bgs}).  This occurs at $T\approx
0.5$ K.

\section{Discussion of results with field applied along $[001]$}\label{dissec001}

Both materials show similar behavior: they are driven into the
$Q=0$ structure by the application of the field and are strongly
hysteretic/history dependent at very low temperature.

When measuring \HoGans, Bl\"ote {\it et al.} noted that below
$T\approx 0.3$ K it appeared that the spin system was apparently
no longer in good thermal contact with the
thermometer~\cite{blote}. Similar effects were observed by Orendac
in more recent heat capacity measurements on Ho$_2$Ti$_2$O$_7$,
below a temperature of $T\approx 0.6$ K the heat capacity became
time dependent~\cite{mopc}. The neutron scattering data is
comparable with this as the hysteresis develops in field scans at
$0.5$ K and below (see Fig.~\ref{d10hk0loops}). The spin dynamics
of \HoTi are slow in this temperature range: no dynamics have been
detected by inelastic neutron scattering~\cite{mjhunpub} and
Matsuhira {\it et al.} found that the susceptibility of \HoTi was
virtually zero below $0.5$ K~\cite{matsu1}. However, Neutron Spin
Echo~\cite{georg} and muon spin relaxation~\cite{whoSR}
measurements have observed some residual spin dynamics at these
low temperatures. The same is also true of
\DyTi~\cite{matsu2,fukazawa,shiftypeak}. Below $T\approx0.5$ K
there must be a cross over from a regime in which the spin
interactions maintain the system in a spin ice state but there are
sufficient dynamics to allow relaxation, to one which is
completely frozen.  This was clearly demonstrated in \HoTi by the
decay of the remnant magnetization into the disordered spin ice
state as the sample was heated above $0.5$ K (see
Fig.~\ref{hk0bgs}).

The presence of magnetization plateaux in the hysteresis loops at
0.05 K is particularly interesting. The temperature (in \HoTins)
and history (in \DyTins) dependence of the hysteresis suggests
that these are due to very slow dynamics.  The system is almost
completely frozen and must be driven into the ordered state by the
field.  An alternative way of looking at this is to ask why a
plateau occurs at a particular value of the magnetization? For
example in \HoTi it is at $\approx 6$~$\mu_\mathrm{B}$
atom$^{-1}$. We propose that because the dynamics are so slow and
the thermal fluctuations so weak, that the system is barely able
to relax. Therefore, the plateau does not represent a partially
ordered structure with a magnetization of $6/10$, or fluctuating
moments contributing an average of $6/10$ of the saturated
magnetization. Instead, it represents a partially ordered spin
ice, in which $6/10$ of the moments have been driven into their
ordered orientation by the field and pinned. The remaining $4/10$
require higher fields to reorient them.

To extend this conjecture we suggest that there must be loops,
chains and clusters of spins that can be oriented relatively
easily. Meanwhile some spins are left on branches, or in regions
not reached by a percolating cluster.  In the absence of thermal
fluctuations these spins become trapped as defects in the $Q=0$
state and only high fields can remove them. This is similar to
work hardening in alloys. When alloys are first worked, defects
can move to release the stress. As the working continues,
remaining defects become entangled or locked together. Eventually
the material no longer responds to the stress and it fails. The
neutron scattering results from \HoTi support this hypothesis. The
system begins in a disordered spin ice state, with diffuse
scattering and no Bragg peaks.  Applying an external field results
in magnetic Bragg scattering from an ordered structure.
Simultaneously the diffuse scattering from short range
correlations decreases.  However, in the intermediate plateau
region both Bragg and diffuse scattering are present, indicating
that the system has regions of both long and short range order.

The postulated liquid-gas critical point was not irrefutably
absent.  The prediction of Harris {\it et al.} can be cast as a
first order phase transition in finite field~\cite{lgcp}.  In both
materials this is observed at 0.05 K, but it is absent above
$0.15$ K in \HoTins. The critical point must lie in this
temperature range, if it does exist.  Using the effective nearest
neighbor coupling strength ($J_{\mathrm{eff}}=1.8$ K~\cite{jeff})
and expected rare earth moment sizes ($10.0$ $\mu_{\mathrm{B}}$
atom$^{-1}$) the critical point is expected at $T_c = 0.29$ K and
$B_c = 0.27$ T for \HoTins. The development of the hysteresis in
\HoTi occurs in the right temperature and field regime but does
not agree in detail with these predictions. The liquid-gas
critical point is a property of a thoroughly equilibrated
simulated system: it may very well be obscured in the real
materials by the lack of dynamics in the temperature range of
interest. Also, the behavior of the dipolar spin ice model in
applied field is largely unknown, it is possible that the critical
point may appear elsewhere in $(B,T)$ space for this model.

\section{Results for application of field along
$[1\bar{1}0]$}\label{ressec110}

\subsection{\DyTi}\label{dy110ressec}

Because two experiments were performed and the path through
$(B,T)$ space became quite complex, we discuss the experiments in
terms of similar field/temperature sections. For data points
between the large $h,h,l$ maps, rocking curves were measured about
the $0,0,l$ axis out to (0,0,10).

\subsubsection{Hysteresis loops at 0.05, 0.3 and 1.9 K}

During the two experiments two hysteresis loops were measured at
both 0.05 (one in each experiment) and 0.3 K (both in the second
experiment), and one at 1.9 K (second experiment).  Information
about the development of spin correlations can be extracted from
the behavior of both Bragg and diffuse scattering.  In zero field
at $0.05$ and $0.3$ K there is diffuse magnetic scattering typical
of the spin ice ground state~\cite{series1}. On applying a
magnetic field this diffuse scattering disappears and is replaced
by diffuse scattering of a new form centered around
antiferromagnetic $Q = X$ positions such as $(0,0,2n + 1)$.
Concomitantly, magnetic Bragg scattering appears at $Q = 0$
positions like $(0,0,2)$ (see Fig.~\ref{tigermaps}).

\begin{figure}
\caption{\label{tigermaps}(Color Online) \DyTi ($B$ //
$[1\bar{1}0]$): Scattering intensity in the $h,h,l$ plane recorded
on PRISMA at 0.05 K in zero field (top), and in 1.5 T (bottom). In
zero field the diffuse scattering is characteristic of the short
range ordered spin ice state~\cite{series1}.  In 1.5 T both Bragg
(at $Q=0$ positions such as $(0,0,2)$) and diffuse (at $Q=X$
positions such as $(0,0,3)$) magnetic scattering is observed
indicating the coexistence of long and short range magnetic order.
Note that the intensity scale is cut off at 2 to reveal the
presence/absence of diffuse scattering, the maximum intensity of a
Bragg peak is much greater.}
\end{figure}

The new diffuse scattering is sharper than the diffuse scattering
observed in the spin ice state and it is distinctively
distributed. To distinguish it from the Bragg Peaks and spin ice
diffuse scattering it will forthwith be referred to as a $Q=X$
{\it feature}. The $Q=X$ features are relatively narrow parallel
to the $h,h,0$ direction, but remain quite wide along $0,0,l$. The
distinctive distribution of scattering in the $Q=X$ features is
clearly seen in the mapping of the scattering plane in 1.5 T
during the first hysteresis loop at 0.05 K (see
Fig.~\ref{tigermaps}). Even in a field of 1.5 T the $Q=X$ features
are not resolution limited.

At 0.3 K the disappearance of the spin ice diffuse scattering and
development of the $Q=X$ features is similar to that described for
0.05 K. The form of the hysteresis loops differ in detail, but
both Bragg scattering and $Q=X$ features were observed. At 1.9 K
only Bragg scattering was observed, the $Q=X$ features did not
appear.

Structure factor calculations show that the $Q=0$ Bragg scattering
is due to the $\alpha$ chains and the $Q=X$ features are due to
the $\beta$ chains (see section~I for definition of
$\alpha$ and $\beta$ chains). The shape of the $Q=X$ features
gives considerable information about the spin correlations induced
in the $\beta$ chains by the application of the field. The
position of the features (i.e. at $Q = X$) and the fact that they
are discrete in reciprocal space shows that there must be
antiferromagnetic correlations between $\beta$ chains. The
anisotropic shape of the features shows that the order is longer
ranged parallel to $h,h,0$ (intrachain correlation length) than
parallel to $0,0,l$ (interchain correlation length). A single
correlated region is therefore a collection of a few chains (with
an interchain correlation length $\zeta=14.2 \pm 0.7$~\AA~this is
1.13 unit cells (in the $1,1,0$ direction) and so 4-5 chains),
each extending $770 \pm 40$~\AA~(intrachain correlation length)
and ordered antiparallel. The actual correlation lengths measured
depend on the history of the sample (these were obtained at 0.05 K
and 1.5 T), but in the most correlated states observed they are
always of this order. We do not know what happens out of the
scattering plane: the $Q=X$ features could be rods or ellipsoidal
features intersecting the scattering plane. Rods would indicate a
two-dimensional ordering of $\beta$ chains in planes parallel to
the scattering plane. Three dimensional ellipsoids would indicate
three dimensional domains.  On the balance of evidence it seems
likely that the interchain correlations are three dimensional as
opposed to two dimensional.

Comparison of the development of the Bragg scattering during
hysteresis loops measured at the same temperature reveals that the
form of the curve depends on the field increments, or ramp rate,
used to measure the loop. At both $0.05$ and $0.3$~K, two
different hysteresis loops have been measured and shown in
Fig.~\ref{dyloops1}.  At each temperature one loop was recorded
with coarse field increments/rapid change in field.  These loops
contain steps or discontinuities in the development of the order
parameter. Subsequent investigations, using finer field increments
generally reduced the number of steps in the loop but those that
remained were more clearly resolved. Also, with fine field
increments, there is a clear difference between 0.05 and 0.3 K.
With fine field increments at 0.3 K the steps disappeared and a
very smooth increase in intensity was observed.  At 1.9 K no
hysteresis was observed.  Again it appears that there is a change
in dynamical regime at low temperature as shown by the
presence/absence of steps at 0.05/0.3 K.  The temperature of the
change is lower in \DyTi than \HoTi (a similar effect was observed
at 0.5 K in \HoTi and discussed in sections~\ref{ho001ressec} and
~\ref{dissec001}) because the coupling strengths are somewhat
lower in \DyTins.  As with the field applied along $[001]$ the
very slow dynamics of spin ice materials in this low temperature
regime are manifested in these observations.

\begin{figure}
\caption{\label{dyloops1}\DyTi ($B$ // $[1\bar{1}0]$): Integrated
intensity of $(0,0,2)$ with field applied along $[1\bar{1}0]$ at
0.05 K and different sweep rates (top) and integrated intensity of
$(0,0,2)$ with field applied along $[1\bar{1}0]$ at 0.05 K and 0.3
K with slow field sweeping (bottom, the second loop at 0.3 K was
not completed). In the first loop the field stepping regime was
much coarser than the second (the loops at 0.05 K were recorded in
separate experiments and the intensities from the first are scaled
to compare with the second). At similar sweep rates there is a
change in dynamical regime between 0.3 and 0.05 K as shown by the
presence/absence of steps at 0.05/0.3 K. Lines are to guide the
eye only, filled symbols indicate the falling field leg.}
\end{figure}

The identical saturation intensity of the $Q=0$ Bragg peaks shows
that the same long range ordered state in the $\alpha$ chains was
realized in the hysteresis loops at 0.05, 0.3 and 1.9 K. Unlike
the Bragg peaks, the maximum intensity observed in the $Q=X$
features was temperature dependent. Comparing hysteresis loops
within the second experiment shows that the $Q = X$ features
become more intense at 0.3 K than at 0.05 K (see
Fig.~\ref{dyamps}).  At 1.9 K the $Q=X$ features do not form.
This shows that for $T > J_{\mathrm{eff}}$ (1.1 K for \DyTins) the
formation of the $\alpha$ chains does not require the formation of
the $\beta$ chains.  However, for $T< J_{\mathrm{eff}}$ the
$\beta$ chains form and the degree of correlation between them
depends on the dynamics.

\begin{figure}
\caption{\label{dyamps} \DyTi ($B$ // $[1\bar{1}0]$): the
amplitude of the $Q=X$ feature at $(0,0,3)$ during hysteresis
loops at 0.05 K and 0.3 K in the second experiment.  The degree of
correlation in the $\beta$ chains is temperature dependent.  At
0.3 K they become more strongly correlated. Lines are to guide the
eye only, filled symbols indicate the falling field leg.}
\end{figure}

\subsubsection{Temperature scans in applied field}

At 1.5~T, after the first field cycle, the temperature was raised
to $\approx$ 1~K. In this temperature range, the $Q=0$ Bragg peaks
were found to be temperature independent.  However the intensity
of the $Q=X$ features passes through a maximum at 0.7 K. Another
temperature scan was performed after the second hysteresis loop at
0.05 K.  This was done in 0.4 T, to begin in the partially ordered
plateau phase.  A more pronounced maximum in the intensity of the
$Q=X$ features was observed at 0.6 K, above which both $Q=0$ and
$Q=X$ scattering decreased.

In \DyTi 1.5 T is sufficient to  maintain the $\alpha$ chains in a
fully polarized state, with no thermal randomization.  The
correlations in the $\beta$ chains increase, but begin to decrease
above a temperature of $\approx 0.7$ K.  The maximum temperature
was not sufficiently high to leave the spin ice regime of \DyTi
and see complete collapse of these features.  The second
temperature scan in 0.4 T began in one of the magnetization step
phases.  Here we see that the rising temperature activated the
development of further correlations in the $\beta$ chains, before
thermal disorder began to reduce the correlations of both $\alpha$
and $\beta$ chains.

\subsection{\HoTi}\label{ho110ressec}

The sample was cooled to $0.05$ K in zero field and a full, 5
quadrant, hysteresis loop was measured (up to 2.5, down to -2.5,
then back up to 2.5 T). The temperature was then raised to $1.2$ K
and then cooled to 0.05 K  whilst maintaining a field of 2.5 T. At
each field/temperature point at least five reflections were
measured using $\omega$-scans, these being $(1,1,0)$, $(0,0,1)$,
$(1,1,\bar{1})$, $(1,1,\bar{3})$ and $(0,0,2)$.  In order to
determine magnetic structures, at certain field-temperature points
an expanded collection of peaks was measured. Where magnetic
structures were refined, crystal structure and extinction
parameters were assumed to be the same as those determined for the
other crystal used with the field applied along $[001]$.  This
assumption resulted in a relatively poor refinement of magnetic
structure in this experiment.  We will therefore discuss most of
the results in this section in terms of integrated intensities
rather than sublattice magnetization.

\subsubsection{Hysteresis loop at 0.05 K}

The entire hysteresis loop is shown for the $(0,0,2)$ peak in
Fig.~\ref{bgloops}.  The scattering at the $Q=X$ positions
$(1,1,0)$ and $(0,0,1)$ is diffuse. During the hysteresis loop
these features never evolved into a peak-shape within the bounds
of the $\omega$-scan. The intensity at the $(0,0,1)$ position is
also shown in Fig.~\ref{bgloops}.

The $(0,0,2)$ Bragg intensity saturates identically on both sides
of the loop.  This implies that the same long range ordered state
is reached in high field, regardless of history and coexisting
short range correlations. After ramping the field and driving the
$\alpha$ spins out of the disordered spin ice state for the first
time (pristine curve) subsequent field sweeps are essentially all
the same. The pristine curve lies outside the subsequent field
cycles, indicating that the spin ice state is harder to magnetize
than the subsequent state composed of $\alpha$ chains.

\begin{figure}
\caption{\label{bgloops} \HoTi ($B$ // $[1\bar{1}0]$): Hysteresis
loop at 0.05 K. The amplitude of the diffuse feature at $(0,0,1)$
was fitted as a flat line (top).  The $(0,0,2)$ peak (bottom) is
directly related to the order in the $\alpha$ chains.  As the
$\alpha$ chains are ordered by the field, correlations are
produced in the $\beta$ chains.  The $\beta$ chains evolve toward
an increasingly correlated state but are only able to do so when
the field produces dynamics in the $\alpha$ chains. Lines are to
guide the eye only, filled symbols indicate the falling field
leg.}
\end{figure}

When the $Q=0$ Bragg peaks appear, so do the $Q=X$ features.
Although the detailed shape of the peak is not accessible in a
single $\omega$-scan, some further details about their form were
established.  It is possible that the features could be part of
extended sheets of scattering. With the field applied along $[1
\bar{1} 0]$ the $\beta$ chains are formed along $[110]$.  A sheet
of  scattering might be expected to extend along $0,0,l$.
$\omega$-scans at an $(0,0,l)$ position should scan across this
sheet.  Our collection of peaks allows us to check this
possibility.  Firstly, the fact that the $(0,0,1)$ feature is
flat, or outside the width of the $\omega$-scan, means that any
such sheet is also rather broad perpendicular to $0,0,l$.  It
would be expected as a flat background contribution at $(0,0,2)$
as well. The form factor of Ho$^{3+}$ will reduce the intensity,
but both positions are quite close to the maximum in the form
factor, so the intensity maximum for $(0,0,1)$ in the hysteresis
loop (600 counts) should also be clearly visible at $(0,0,2)$.
Likewise the presence of $(1,1,0)$ as a flat feature would imply
another sheet extending along $1,1,l$ which would be visible in
the background of $(1,1,\bar{1})$. However, the backgrounds of
both Bragg peaks are field-independent.  This rules out the
possibility of extended sheets of scattering parallel to either
$0,0,l$ or $h,h,0$ and suggests that the $Q=X$ features are
localized, as in \DyTins.

The evolution of the $Q=X$ features throughout the hysteresis loop
is very interesting (see  Fig.~\ref{bgloops}).  In general the
intensity increases from the beginning to the end of the field
cycling (the full 5 quadrants). However, it does so in a
discontinuous, ratchet-like manner.  When the intensity of the
$Q=0$ peaks is saturated, there is no change of intensity at the
$Q=X$ positions. Only when the hysteresis loop passes through a
region where there are field induced changes in the $Q=0$
intensity do changes occur in the $Q=X$ features.

Despite the $\beta$ chains being decoupled from the field, it is
clear that at this temperature the field regulates the dynamics of
both $\alpha$ and $\beta$ chains. As discussed in
section~\ref{dissec001}, it seems certain that a spin ice is
frozen as regards spin flips in zero field: any significant spin
flip dynamics must therefore be produced by the action of the
field. During the hysteresis loop it can be seen that not only
does the field reorient the $\alpha$ chains, but in doing so
propagates dynamics into the $\beta$ chains. During these parts of
the hysteresis loop the degree of correlation increases in the
$\beta$ chains. When the $Q=0$ structure does not change, neither
does the $Q=X$.

Two possible magnetic structures, with $Q=0$ and $Q=X$ were
discussed in the earlier work of Harris {\it et al.}~\cite{prl1}.
These structures were assumed to involve all four spins of the
tetrahedral basis, which in light of these new results, is clearly
not appropriate. We therefore used a magnetic structure with zero
moments on the $\beta$ chains and ferromagnetic $\alpha$ chains to
fit the magnetic Bragg peaks (ignoring the diffuse scattering).
The fitted moment ($6.5 \pm 0.3~\mu_{\mathrm{B}}$ atom$^{-1}$) was
less than the expected 10.0 $\mu_{\mathrm{B}}$ atom$^{-1}$ for
Ho$^{3+}$, consistent with the findings of Harris {\it et
al.}~\cite{prl1} who refined the moment on all four spins,
assuming a fully ordered spin system. We believe the observed
reduced moment is an artifact of a poor correction for extinction.

\subsubsection{Temperature Scans in applied field}

After the hysteresis cycle returned to 2.5 T the system was warmed
to 1.2~K and then cooled back to 0.05 K in a field of 2.5 T.  An
enormous gain in intensity is seen at the $Q=X$ positions, in
particular $(0,0,1)$, above 0.3~K. The broad feature outside the
width of the $\omega$-scan at low temperature, develops into a
quite sharp peak. The correlations between the $\beta$ chains are
definitely three dimensional at this stage. However, these
features are not resolution limited and do not have the form of a
Bragg peak in the detector volume. The temperature dependence is
shown in Fig.~\ref{bgtscan}. There is a tiny increase in the $Q=0$
Bragg peak intensity during the temperature cycle.

Clearly the rising temperature activates the dynamics of the
$\beta$ chains and because $T<J_{\mathrm{eff}}$ the restoration of
dynamics increases correlations rather than destroying them.
During the temperature rise $\chi_\mathbf{Q}$ begins to fall again
above $T\approx 1$ K.  This is roughly consistent with the peak in
the susceptibility seen at around 1 K~\cite{matsu1}. Lowering the
temperature again then freezes these correlations.  Activating the
dynamics of the $\beta$ chains also allows a further annealing of
the $\alpha$ chains causing the slight increase in Bragg
intensity.

\begin{figure}
\caption{\label{bgtscan} \HoTi ($B$ // $[1\bar{1}0]$): temperature
scan in a field of 2.5 T. The amplitude of the diffuse feature at
$(0,0,1)$ was fitted as a flat line (0.05-0.3 K), single gaussian
(0.3-0.6 K) or triangular peak (0.6-1.2 K and all points on the
falling $T$ scan).  For $T < J_{\mathrm{eff}}$ increasing the
temperature activates the dynamics of the $\beta$ chains and also
causes their level of correlation to increase.  Once $T>
J_{\mathrm{eff}}$ the activated dynamics begin to destroy the
correlations, this is evident in the down turn of $(0,0,1)$
intensity at 1.1 K. Lines are to guide the eye only, filled
symbols indicate the falling temperature leg.}
\end{figure}

\section{Discussion of results with field applied along
$[1\bar{1}0]$}\label{dissec110}

The field-dependent properties of \DyTi and \HoTi with field
applied along $[1\bar{1}0]$ show many generic features. In this
section we discuss these and the conclusions which can generally
be drawn for spin ice materials with a field along $[1\bar{1}0]$.

The spin system is separated into two sub-systems, the $\alpha$
and $\beta$ chains. This is inferred from the coexistence of long
and short range order, the field induced short range order, and
the contrasting temperature behavior of the $Q=0$ and $Q=X$
scattering. When the field pins the $\alpha$ spins  into a long
range ordered structure, the manifold of degenerate states is
restricted considerably and so this also causes the $\beta$ spins
to become more correlated.  The correlations in the $\beta$ chains
cause the scattering at the $Q = X$ positions. Although the
evolution of the ordered $Q=0$ phase is not identical in all
hysteresis loops the $\alpha$ chains  always reach the same fully
polarized state, independent of the state of the $\beta$ chains.

In both \DyTi and \HoTi the scattering at the $Q=X$ positions
appears immediately, on application of a field, as a localized
feature which then sharpens. This means that the interchain
correlations begin to form as soon as the chains appear.  This
observation may be compared to those of Hiroi {\it et al.}  who
made a calorimetric study of \DyTi in  a field applied along
$[110]$~\cite{hiroi}. On the basis of these results they argued
that the $\beta$ chains behaved effectively as an assembly of one
dimensional ferromagnetic Ising chains embedded in a three
dimensional crystal.  We have directly measured the intrachain
correlation length:  at 1.5 T and 0.05 K it is of order $800$~\AA~
for \DyTins. The very long but finite intrachain correlation
length is indeed a characteristic property of the one dimensional
Ising ferromagnet.  Our results clearly show that there are also
correlations between the chains.  At 1.5 T and 0.05 K the
interchain correlation length is of order $15$~\AA~for \DyTins.
The actual value of the correlation length depends on the history
of the sample.  The quoted values were at 0.05 K, longer
correlation lengths are observed in the hysteresis loops at 0.3 K.

Hiroi {\it et al.} pointed out that these chains are arranged in a
triangular array, which frustrates the coupling between them,
preventing the development of true long range order. The
triangular arrangement is not perfectly equilateral, so that in
theory a transition to $Q=X$ order is observed in the $\beta$
chains.  Our results also show clearly that the spin system does
not achieve the long range order predicted at this point in
$(B,T)$ space~\cite{yoshida,ruff}.

Indirectly these experiments have probed the dynamical processes
operating in the spin ice materials in this very low temperature
regime.  The unusual situation of having the $\beta$ chains
perpendicular to the field and yet coupled to it by their
interaction with the $\alpha$ chains gives particular insight into
the dynamical regimes.  These measurements show that there are
dynamical processes in these materials at this extremely low
temperature which have not yet been explained or detected in
microscopic detail. The dynamics are extremely slow as these
experiments operate at the longest timescales and yet equilibrium
is not maintained. Because we are below $T = J_{\mathrm{eff}}$ the
density of single spin flips must be exponentially small so they
must be correlated dynamical processes.

At 0.05 K both materials appear to be frozen: the correlations in
the $\beta$ chains only evolve when dynamics are restored by the
field, via the $\alpha$ chains.  As the temperature is raised
dynamics are increased.  The boundary of the fully frozen regime
was not clearly established.  In all the experiments the histories
followed placed the magnet in a strong field at very low
temperature, with relatively disordered $\beta$ chains (i.e. there
was considerable potential for further order to develop in the
$\beta$ chains, as occurred subsequently).  In both materials this
lead to the observation that the $\beta$ chains can be freed from
the $\alpha$ chains and exhibit their own independent behavior.
This was most clearly shown in \HoTi where the rising temperature
lead to a startling increase in sharpness and intensity of the
$Q=X$ features.  Similar observations were made in \DyTi where the
degree of correlation achieved in the $\beta$ chains was greater
in a hysteresis loop at 0.3 K than at 0.05 K.  As far as they are
known, the measured dynamics of spin ices are consistent with
this~\cite{matsu1,matsu2,fukazawa,georg}.  Dynamical processes
appear to persist to lower temperatures in \DyTi which is
consistent with susceptibility measurements.

During all the experiments the spin ices appear to seek a long
range ordered groundstate, but never realize one.  This is shown
by the ever increasing degree of correlation in the $\beta$ chains
when dynamics are restored and the $\alpha$ chains are controlled
by the field.  The increasing correlations are characteristic of
the formation of the $Q=X$ structure.  The transverse ordering
must be controlled by long range interactions, as the chains are
not coupled by nearest neighbor interactions. It remains unclear
whether the equilibrium state is the fully ordered $Q=X$ structure
predicted by Yoshida ~\cite{yoshida} rather than the
extended-range correlated state actually observed. The work of
Harris {\it et al.} also should be examined~\cite{prl1}.  They
concluded that the $Q=X$ structure was a high temperature
modification of the $Q=0$ structure.  Here it is shown that
because of the non-equilibrium nature of the low temperature
regime, the magnetic structure observed is very much controlled by
the dynamics and history of the system.  The question of the
existence and spontaneous selection of a groundstate in spin ice
remains open.  It is surprising that after these extensive travels
in $(B,T)$ space, the materials seem to be arrested on the brink
of order. Even at $T\approx 1$ K where the susceptibility is large
and there are considerable dynamics available in \HoTi it is
unable to locate an ordered groundstate.  It would be interesting
to establish how the material becomes jammed in this state and why
the dynamics are unable to extend the correlations to long range
order.

At low temperatures ($T<0.5$ K) the magnetization of the $\alpha$
chains depends on the rate of sweeping the field.  In all
hysteresis loops observed in this regime (except one with a slow
field sweep rate in \DyTins) the $\alpha$ chain magnetization
develops discontinuously through a series of jumps.  In \HoTins,
where repeated cycles were made, the pristine order parameter
curve lies outside of the main hysteresis loop which shows that
the zero field cooled spin ice state is harder to magnetize than
the field-cycled state.  Neutron scattering cannot identify if the
order parameter jumps involve symmetry sustaining first order
phase transitions of the type discussed in Ref.~\onlinecite{lgcp},
or they are a purely kinetic effect, involving the abrupt growth
of macroscopic domains. If the latter, the behavior of spin ice in
this regard is reminiscent of random anisotropy Ising magnets such
as amorphous Dy$_{0.41}$Cu$_{0.59}$~\cite{coey} where similar
magnetization jumps (like a giant Barkhausen noise) are observed.

Field-induced magnetization steps with sweep rate dependence have
also been observed in the systems Gd$_5$Ge$_4$ and
La$_{0.6}$Ca$_{0.4}$Mn$_{0.96}$Ga$_{0.04}$O$_3$~\cite{dmckpmartensitic}.
Both those systems have a field induced crossover from an
antiferromagnetic to a ferromagnetic groundstate, accompanied by a
structural transition. The steps were attributed to the burst-like
growth of ferromagnetic regions in an antiferromagnetic matrix.
The sweep rate dependence is believed to be due to the difficulty
of accommodating the structural intergrowths at faster field sweep
rates.  Such behavior can also be attributed to the existence of
two competing groundstates~\cite{dimitricom}.  Here we have no
structural phase transition but there are two distinctive states
at either end of the field sweep, the entropically favored spin
ice state in zero field and the energetically favored $Q=X$ state
in high field.  The situation with the field applied along $[001]$
is analogous but with a different field induced groundstate.  We
suggest that these transitions might be viewed in the same way.
Long range order grows in bursts out of the disordered spin
system.  In the martensitic transitions the bursts are due to the
fact that system is well below the energy scale of the
electrostatic interactions which maintain the structure and so the
dynamics of the structural rearrangement are extremely slow.  Here
the system is well below the energy scale of spin-spin
interactions and the establishment of the ice rules regime
severely restricts dynamics by preventing simple spin flips.
Consequently all the transitions are being controlled by the
balance of spin-spin interactions and magnetic field, with the
small effective temperature promoting metastability.

\section{Conclusions} \label{consec}

We have investigated the field-induced order in the spin ices
\HoTi and \DyTins. The degeneracy removal scheme depends upon the
direction along which the field is applied.  As expected, the
application of a field along [001] leads to the selection of a
long range ordered state of the $Q=0$ type. Application of the
field along $[1\bar{1}0]$ leads to the separation of the system
into $\alpha$ and $\beta$ chains.  It has been shown that the
$\beta$ chains have a long but finite intrachain correlation
length and only a very short interchain correlation length, and
that the degree of correlation is history-dependent.  A pathway to
spontaneous selection of a long range ordered groundstate was not
found, despite partial constraint of the system by the field.

A prevalent feature of the experiments was the strong history
dependence, usually manifested as hysteresis on cycling the field.
It is concluded that as the experiments operate on such a long
timescale, the dynamics of a spin ice are extremely slow in this
temperature range.  These slow dynamics mean that metastable
states and coexisting long and short range order could be observed
in both orientations.  It was also shown that there is a change of
dynamical regime from one in which the spin ice correlations are
established but dynamical processes operate, to one in which the
spin ice is much more strongly frozen.  For measurements such as
these this occurs at $T\approx 0.5$ K for \HoTi and $T\approx 0.3$
K for \DyTins.  The determination of the microscopic details of
these dynamics remains a difficult problem.

\section{Acknowledgements}

We gladly acknowledge the dedication of the sample environment
teams at ISIS (R. Down and D. Bates) and the ILL (J.L. Ragazzoni)
and members of the theoretical group at the University of Waterloo
(M. J. P. Gingras, B. C. den Hertog, M. Enjalran, R. G. Melko and
T. Yavors'kii) for discussions throughout this work.  We
acknowledge the EPSRC for funding of beamtime and a studentship
(TF). TF thanks Andrew Harrison for an extended stay in Edinburgh
while this work was written up.


\begin{thebibliography}{52}
\bibitem[{sph()}]{spheric}
\bibinfo{note}{The crystal was tumbled in a turbulent airflow, inside a boron
  carbide lined chamber. After several days they are almost spherical, with
  vestiges of the original shiny faces which aid alignment.}
\bibitem[{rfa()}]{rfactor}
\bibinfo{note}{$R_3$ is the weighted $R$-factor $\sum_i
  W_i(I_i^\mathrm{obs}-I_i^\mathrm{calc})/W_i(I_i^\mathrm{obs})$, where
  $W_i=1/\sigma_i^2$ is the weight of observation $i$. If $R_3<10\%$ this
  indicates an acceptable model in conventional crystallographic terms.}
\bibitem[{jef()}]{jeff}
\bibinfo{note}{$J_{\mathrm{nn}}$ is the effective exchange interaction energy
  $J_{\mathrm{eff}} = J_{\mathrm{nn}} + D_{\mathrm{nn}}$~\cite{byron}. For
  Ho$_2$Ti$_2$O$_7$ $J_{\mathrm{eff}} = 1.8$ K and for Dy$_2$Ti$_2$O$_7$
  $J_{\mathrm{eff}} = 1.1$ K.}
\end{thebibliography}
\end{document}